\documentclass[letter]{aa} 
\usepackage{graphicx}
\usepackage{txfonts}
\newcommand{\nc}{\newcommand}
\nc{\Teff}{$T_{\rm eff}$\,}
\nc{\logg}{log~$g$~}
\nc{\CCC}{${\rm ^{13}C}$}
\nc{\CC}{${\rm ^{12}C}$}
\nc{\Crat}{\CC/\CCC}
\nc{\mic}{$\mu$m}
\begin{document}
   \title{The extra-mixing  efficiency in very low
   metallicity RGB stars\thanks{Based on observations collected
with the VLT/ISAAC instrument at Paranal Observatory, ESO (Chile) -
Programme 75.D-0228A}}


   \author{A. Recio-Blanco
          \and
          P. de Laverny
          }


   \institute{Dpt. Cassiop\'ee, Observatoire de la C\^ote d'Azur, 
CNRS/UMR 6202, BP4229, 06304 Nice cedex4, France\\
              \email{[arecio;laverny]@obs-nice.fr}
             }

   \date{Received 2006; accepted }

 
  \abstract
   {}
   {After the first dredge-up, low-mass Red Giant Branch (RGB) stars
   experience
     an extra-mixing episode that strongly affects the chemical abundances
    on their surface. This mixing occurs at the bump in the 
    luminosity function.  In this
    Letter we describe the efficiency of the extra-mixing in  RGB stars found 
    in very metal-poor globular clusters (GC).
   }
   {The VLT/ISAAC spectra of twenty stars located 
    between the bump and the tip of the RGB in four GCs with metallicities
    between [Fe/H]=-1.2 and -2.5~dex were collected.
    The carbon isotopic ratios on their  surface were derived from
    the second overtone ($\Delta$v=2) bands of the CO molecule at
    2.3\mic \ with the spectral synthesis method.  
   }
   {It is found that the carbon isotopic ratios of very metal-poor
    GC stars always reach the equilibrium value of the CNO cycle almost
    immediately above the bump in the 
    luminosity function. No additional mixing episode at brighter luminosities
    and no variations 
    with the clusters' metallicity were detected. The extra-mixing is 
    therefore found to be very 
    efficient in metal-poor low-mass RGB stars, in very good agreement with
    theoretical expectations. }
   {}


   \maketitle
%

\section{Introduction}

During their  ascent   of the   Red   Giant Branch   (RGB),  low-mass  
stars undergo mixing  processes that bring  freshly  synthetised  nuclides
to their surface.   In  standard  stellar evolution   
theory (Iben, \cite{Iben}),    
chemical changes in the   surface
are only expected  to be caused by  convective dilution during the first
dredge-up (1DUP)
occurring at the base of the RGB: the  convective  envelope   deepens and 
enters  regions   already
processed  by the central H-burning phase,  thereby  altering the surface
abundances. 
This process leads to a  decrease in the carbon isotopic ratio
(\Crat) from the main sequence value ($\sim$90 in the solar case)
to post 1DUP
values around  20-25 (Charbonnel \cite{Charb4}  or 
Denissenkov \& Herwig \cite{Den}).   In
addition, the carbon abundance drops, while the
nitrogen one increases.  Oxygen and all
heavier element abundances,  however, would remain unchanged.
According to the standard  scenario, the surface abundances after
the 1DUP   remain unaltered, as  the   convective envelope
slowly withdraws outward in mass during the final stages of the RGB evolution.

This canonical picture is, however, challenged by an
increasing  amount  of observational   data demonstrating  its limited
validity; see, for instance, Brown \& Wallerstein \cite{Brown} and, more 
recently, Smith et al. \cite{smith2} and Gratton et al. \cite{Grat}).
In fact, in most of  the field and globular
cluster (GC) low-mass evolved stars, the observed conversion of \CC \ to \CCC \
and $^{14}$N greatly exceeds the levels expected from standard stellar
models.
More recently, Shetrone (\cite{Shet}) derived \Crat \ ratios in 
low-metallicity GC RGB stars ([Fe/H] $>$ -1.2) and found very low values,  
almost reaching the   near-equilibrium value  of the  CNO cycle
(\Crat $\sim$ 3.5).
This very clear evidence  requires a non canonical  mixing
(named {\it extra-mixing}) between the shell and  the bottom of the
convective  envelope (see for instance Charbonnel \cite{Charb4}, and
Weiss \cite{Weiss}).  
Most  of the  observations
supporting this  evolutionary scenario  indicate that extra-mixing
begins at the RGB luminosity-function bump.
This bump appears when the narrow burning H-shell reaches the sharp  
chemical discontinuity in the
H-distribution profile caused by the deep penetration of the
convective envelope. A drop in 
the stellar luminosity then occurs,
revealed by a bump in the RGB luminosity distribution.
This bump, predicted by Iben (\cite{Iben2}) 
and detected by King et al. (\cite{King}), occurs at 
brighter luminosities for more metal-poor GCs.

As already proposed  by
Charbonnel (\cite{Charb}, \cite{Charb5}), observations   strongly
suggest that prior to the bump, 
the  mean molecular weight  gradient
created during  the 1DUP acts as  a barrier to  any mixing
below  the  convective envelope.   After the bump, 
the gradient of
molecular weight above the H-burning shell is much lower, allowing the
extra-mixing to act.
Now, the extra-mixing scenario is  well admitted,  
but its exact physical nature and  efficiency  are still objects of
debate (see, for instance, Palacios et al. \cite{Ana}). 
Furthermore, the 
extra-mixing efficiency at very low metal content is still poorly
known, as is its metallicity dependence. We therefore
present here carbon isotopic ratios
of RGB stars in GC ten times metal-poorer (down to [Fe/H] $\sim$ -2.5) than
presented in Shetrone (\cite{Shet}). They were
derived from low-resolution spectra
around the second
overtone bands of the CO molecule at 2.3\mic.
The selection of the targets, the observations, and their reduction 
are presented in Sect.~2.
Sect.~3 is devoted to the derivation of the \Crat \ ratios
and, we discuss the extra-mixing efficiency in such
metal-poor stars in Sect.~4.

\section{Target selection and observations}
We selected three GCs with [Fe/H] ratios about ten
times lower than in previous works : 
NGC~6397, M30 (NGC~7099), and 
M15 (NGC~7078) with [Fe/H]=-2.1, -2.3, and -2.45~dex, respectively. 
For the purpose of a comparison with Shetrone (\cite{Shet}),
we also selected M4 (NGC~6121 with [Fe/H]=-1.2~dex).
For all these clusters (see Table~1), we adopted the metallicity scale
of Kraft \& Ivans (\cite{Kraft}). 
The observed $K_s$-magnitudes of the M15 and M4 bumps were taken
from Cho \& Lee (\cite{Cho}; Ferraro et al \cite{Ferr}
and Valenti et al. \cite{Val} report
a consistent value for M15). For M30 and NGC~6397, we estimated
their $K_s$-bump magnitude from the relation derived by Valenti et al.
(\cite{Val}), from the 
photometric systems transformation of Carpenter (2001), 
and from distance moduli and extinctions in the Harris catalogue
(\cite{Harris}, see next section for more details). 
This relation was checked for a
very good agreement with the observed $K_s$-bump magnitudes
of M15 and M4.
The adopted $K_s$-magnitudes of the GC bump are reported in Table~1.
Typical uncertainties for $M_K$-bump magnitudes are $\pm$0.1mag.
(Valenti et al. \cite{Val}).
Taking into account errors on distance moduli and extinctions,
the total error for the calculated $K_s$-bump magnitudes 
is of the order of 0.15mag. It is around 0.1mag. for the observed
ones (Cho \& Lee \cite{Cho}).

Target selection was based on the photometry by
Rosenberg et al. (\cite{Ros1}) for M4 and NGC~6397, Rosenberg et al.(\cite{Ros2})
for M15, and Momany et al. (\cite{Yazan}) for M30. 
The colour-magnitude diagrams of each GC
were used to select stars located between the bump (or slightly
less luminous) up to the tip of the RGB.
The observed stars are listed in Table~1 with the naming
convention of their corresponding photometry. 
Some fainter targets of the sample were finally not included in this
work due to an unclear identification of their CO bands.
We also checked the radial velocity of each target, derived from
the collected spectra, in order to confirm their GC membership.

The observations were collected during the nights July 25-27, 2005, 
with the ISAAC instrument on the VLT/ANTU (ESO, Chile).
We used a slit width of 1", leading to a spectral resolution
of about 3\,000 over the domain 2.275-2.39\mic.
Objects were observed at two positions along the slit to remove
the sky contribution (nodding technique). In most of the cases,
the slit orientation was chosen in order to observe 
two objects simultaneously. Their position on the slit
and the nodding/jitter parameters were defined to 
avoid spectra overlap. The number of nodding cycles varied from 3 to 5
and the detector integration time from 3.5 to 300~sec,
depending on the star magnitude. The total exposure
times (without overheads) ranged between one minute for the brightest stars
at the tip of M4 and NGC~6397, down to one hour for the faintest targets
above the bump of M30 and M15.

To remove the lines of the Earth's atmosphere,
several telluric standards were observed at the same airmass and
with the same instrumental configuration as the science targets,
soon before or after them. The selected standards were
featureless hot OB spectral type stars.
The spectra were reduced with the ISAAC pipeline
offered by ESO, and
standard IRAF procedures were used for spectra extraction.
Each target spectrum was then divided by the spectrum
of its corresponding telluric standard.
In practice, several telluric spectra were tried
in order to remove the telluric lines as fully as possible.
Finally, the observed spectra were normalised 
as proposed by Shetrone (\cite{Shet}). 
We first compute a synthetic spectrum with parameters
and chemical abundances (see Sect.~3) as close as possible to the observed one.
Then, the residual
of the division of the observed spectrum by the synthetic one
is fitted by a low-order spline, which is then used to
normalise the observed spectrum. This procedure was performed iteratively 
until the best synthetic spectrum was found.

\section{Determination of the carbon isotopic ratios}
The carbon isotopic ratios were derived with the 
spectral synthesis method. 
Theoretical spectra  were
computed  with the  turbospectrum code
(Alvarez \& Plez \cite{Plez}, and further improvements by  Plez).
Model atmospheres were interpolated in a grid of new-generation 
MARCS models (Gustafsson et al. \cite{Gus1}, \cite{Gus2}). 
These models are in LTE, spherical geometry and with an enhancement
of the $\alpha$-elements typical of such low-metallic environments
([$\alpha$/Fe] = +0.4).
The CNO abundances of the Sun revised by Asplund et 
al. (\cite{martin}) were adopted, and Solar abundances of 
other chemical species are
from Grevesse \& Sauval (\cite{GS}). 
The adopted line list consists in atomic lines from the
VALD database (Kupka et al., \cite{Vald}) and in CO lines,
including its  different isotopes, from 
(Goorvitch \& Chackerian, \cite{CO}). Other molecular lines  were
found to be invisible in the studied domain at such low metallicities. 
Finally, synthetic spectra were broadened by convolution with a 
Gaussian to match the observed line widths.
This line list was checked by fitting the high-resolution IR spectrum
of Arcturus (Hinkle et al., \cite{hinkle}) with stellar parameters and 
abundances from Peterson et al. (\cite{Peter}).
We found \Crat \ = 6$\pm$1 for the Arcturus spectrum degraded to our adopted
ISAAC resolution, in very good agreement with previous
determinations published since Griffin (\cite{griffin}).

For each target star, the effective temperature was derived from
the ($V-K$) and ($J-K$) colour indices and the Alonso scale
(\cite{Alonso1}, \cite{Alonso2}),
using the transformation equations between the different
photometric systems from Carpenter (2001) and Bessel \& Brett (1988). 
The IR magnitudes are from the 2MASS catalogue by
Skrutskie et al. (\cite{2mass}).
We estimated the extinction in $K$ 
from the GC foreground reddening reported by Harris (\cite{Harris}, 
revised in 2003) and from the relations given in Cardelli et al. (\cite{Card}).
For M4, differential reddening was estimated from Cudworth \& Reed
(\cite{Cud}) but, in the effective temperature determination of its
members, we favoured the estimate derived from the ($J-K$) colour,
as it is less affected by the reddening.
The surface gravity was calculated from the estimated effective 
temperatures, distance moduli (Harris catalogue), and 
isochrones from Pietrinferni et al. (\cite{Santi}).
Finally, we estimated the micro-turbulent velocity from
the empirical relation for metal-poor giant stars 
derived by Pilachowski et al. (\cite{Pila}).

Then, the carbon isotopic ratio of the target 
stars was derived following a 
procedure very close to the one adopted by Shetrone (\cite{Shet}).
The metal abundance of each star was assumed to be the same
as the one of its parent cluster (in the Kraft \& Ivans scale, \cite{Kraft}; 
see Table~1),
and the oxygen abundance was 
adjusted from the $\alpha$-enhancement of these GCs.
We then varied the carbon abundance to match the \CC O(4-2) band at 2.35\mic.
The \Crat \ ratio was finally determined by fitting the 
\CCC O(2-0) band at 2.345\mic. We recall that the observed continuum
was corrected when necessary with the new synthetic spectra
(see Sect.~2) until a stable solution was achieved.
We found that
these two CO bands  give the most reliable solution.
It was indeed impossible to remove 
the telluric contamination accurately enough 
around the other \CCC O band available at 
2.37\mic. Finally, 
the adopted procedure was found to be insensitive to uncertainties 
in the stellar
parameters, since both \CC O 
and \CCC O features vary with them in a similar way.
Furthermore, because the \CC O band are not saturated in these
very-low metallicity atmospheres, the uncertainty
on the micro-turbulent velocity was found  not to affect
the results.
Thus, once the \CC O band is fitted by assuming a given carbon
abundance (and other stellar parameters and chemical abundances), 
only the \Crat \ ratio
can be varied to fit the \CCC O feature. 

The derived carbon isotopic
ratios are given in Table~1.
The corresponding errors are completely
dominated by (i) the signal-to-noise ratio of the observed stellar
and telluric spectra, (ii) the removal of the telluric absorptions,
and (iii) the normalisation of the observed spectrum (these last two
points being the dominant source of error).
The error bars reported in Table~1 have been estimated
by varying the \Crat \ ratio,
taking all these uncertainties into account and, in particular,
the continuum location around
the \CC O and \CCC O bands. For low \Crat \ values, errors are found
to be much smaller due to the greater strength of the \CCC O band. 
For higher \Crat, this band can be very weak, leading
to much larger error bars.

\begin{table}[t]
\caption{Target properties and derived carbon isotopic ratios. 
The adopted metallicities (in dex)
of each GC are in the Kraft \& Ivans scale
(\cite{Kraft}). See text for references on
$V$, $K_s$, and $K_s^{\rm Bump}$-magnitudes and discussion of their errors.}           
\label{Tab} 
\centering   
\renewcommand{\footnoterule}{} 
\begin{tabular}{c c c r c c l}     
\hline\hline  
& & & & & & \\      
\multicolumn{3}{c}{Globular Cluster} & Star & $V$ & ($K_s$-$K_s^{\rm Bump}$) & \Crat\\
Name & [Fe/H] &  $K_s^{\rm Bump}$   &      &     &                     &      \\
& & & & & & \\ 
\hline                      
& & & & & & \\
M4 & -1.2 & 10.0 & 304 & 11.5 &~~2.8 & ~~4$_{-1}^{+1}$ \\
   &      &      & 113 & 12.5 &~~1.3 & ~~5$_{-1}^{+2}$\\ 
   &      &      & 213 & 12.9 &~~0.7 & ~~5$_{-1}^{+2}$\\
   &      &      & 456 & 13.5 &~~0.1 & ~~7$_{-2}^{+5}$\\ 
   &      &      & 148 & 13.4 &~~0.0 & ~~8$_{-3}^{+5}$\\
   &      &      &  53 & 13.5 &~-0.1 & ~~6$_{-1}^{+5}$\\ 
   &      &      & 152 & 13.7 &~-0.4 & $>$10\\
& & & & & & \\   
\hline 
& & & & & & \\ 
NGC~6397 & -2.1 & 9.7  & 1114 & 9.9 & ~~3.4 & ~~4$_{-1}^{+1}$ \\
        &      &      & 224 & 10.7 & ~~2.2 & ~~5$_{-1}^{+2}$\\
        &      &      & 110 & 11.5 & ~~1.2 & ~~5$_{-1}^{+3}$\\
        &      &      & 429 & 11.5 & ~~1.2 & ~~4$_{-1}^{+3}$\\
        &      &      & 183 & 11.8 & ~~0.7 & $>$5 \\
& & & & & & \\
\hline                        
& & & & & & \\
M30 & -2.3 & 12.2 & 7917 & 12.1 & ~~3.4 & ~~6$_{-3}^{+4}$\\
    &      &      & 3998 & 12.1 & ~~3.3 & ~~5$_{-2}^{+3}$\\
    &      &      & 7640 & 12.6 & ~~2.4 & ~~7$_{-3}^{+6}$\\
    &      &      & 3711 & 13.0 & ~~1.9 & ~~7$_{-4}^{+8}$\\
    &      &      & 7927 & 13.2 & ~~1.7 & 10$_{-5}^{+15}$\\
& & & & & & \\
\hline 
& & & & & & \\
M15 & -2.45 & 12.9& 1662 & 12.8 & ~~3.5 & ~~4$_{-1}^{+2}$\\
    &       &     &   58 & 13.3 & ~~2.6 & ~~4$_{-1}^{+2}$\\
    &       &     &  665 & 13.5 & ~~2.3 & ~~4$_{-1}^{+2}$\\
& & & & & & \\
\hline 
\end{tabular}
\end{table}

\begin{figure*}
\begin{center}
  \includegraphics[width=13cm]{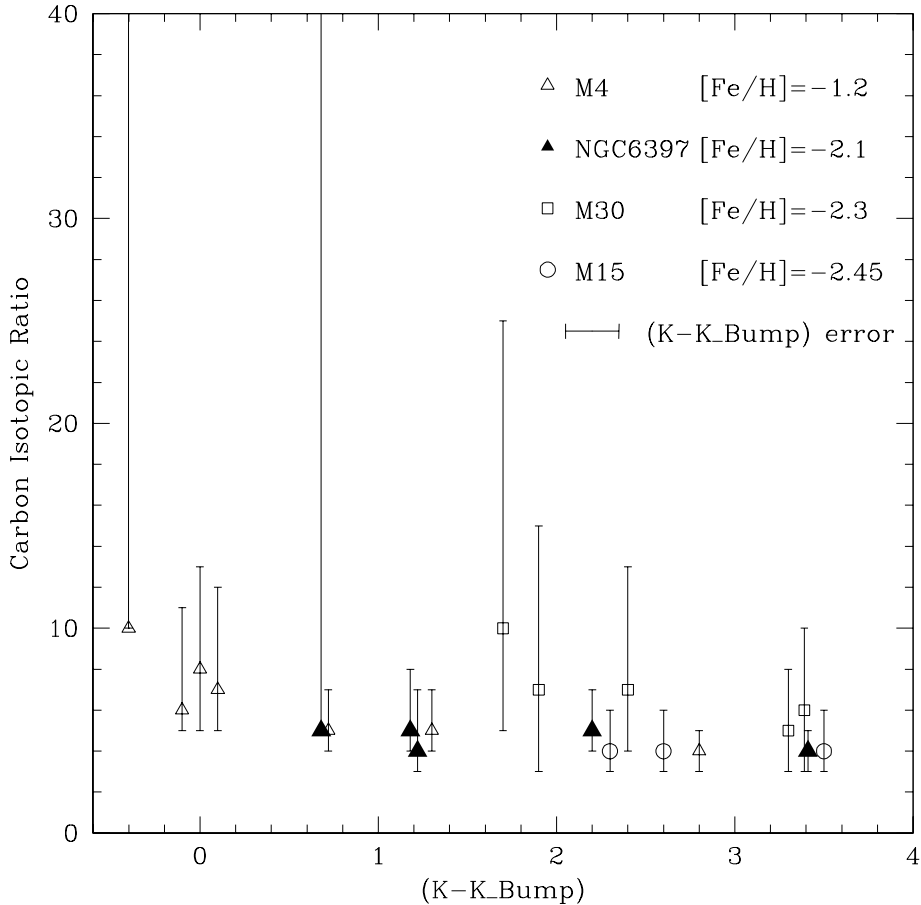}
  \caption{Evolution of the carbon isotopic ratio with respect
to the stellar $K_s$-magnitude above the bump in the RGB luminosity
function of each globular cluster. The adopted $K_s$-bump magnitudes
are reported in Table~1.
The maximum error on  ($K_s-K_s^{\rm Bump}$) is shown in the upper
right corner and, for a given cluster, the uncertainty on the
relative stellar luminosities with respect to their bump 
is $\pm0.04$~mag. (see text for more details).
The reported metallicity (in dex)
of each GC is from Kraft \& Ivans (\cite{Kraft}).
For two stars (one in M4 and one in NGC~6397), we have only been able
to derive upper limits for their carbon isotopic ratios. These limits
are represented
by a symbol located at the lower end of their error bar.}
\end{center}
\end{figure*}

\section{Extra-mixing efficiency in very metal-poor stars}
Figure~1 presents our derived carbon isotopic ratios with respect to the 
evolutionary status of the targets above the luminosity function bump.
Comparing the stellar $K_s$-magnitude with the $K_s$-magnitude
of the bump as in Fig.~1 has the advantage of avoiding uncertainties in
distance moduli and interstellar absorptions. Moreover, for the case of M4,
this procedure diminishes any uncertainties due to diffential reddening.
Errors in ($K_s-K_s^{\rm Bump}$) are therefore dominated by the
uncertainty on $K_s^{\rm Bump}$ and are close to $\pm$0.15~mag in the
worst cases (see Sect.~2).
Of course, for a given cluster, the uncertainty on the
relative positions of its members with respect to the bump luminosity 
is much fainter, since the 2MASS photometric error is about
$\Delta K_s \simeq 0.04$~mag.

It can be seen that the carbon isotopic ratios
are always found to be very low and close to or equal
to the theoretical equilibrium value (within error bars),
even immediately above the bump.
In the case of NGC~6397 and M15, the \Crat \ ratio is indeed
found to always be smaller
than 5 for a wide range of luminosities on the upper RGB,
starting just above the bump.
For M30, slightly higher values (around 6 in average)
are found, but they are still compatible with the CNO equilibrium value
and the extra-mixing scenario.
We therefore do not detect any
variation with metallicity: the extra-mixing
appears to be very efficient whatever the value of [Fe/H] is,
leading to a large decrease in the carbon isotopic ratio
down to the equilibrium value.
Unfortunately, we were unable to derive carbon isotopic ratios
of stars fainter than the bump for these very metal-poor GCs 
in order to precisely define where the extra-mixing starts
and what its exact duration is. This would require much larger SNR
spectra with higher spectral resolution.
Finally, there is a group of three M4 stars
close to the bump but with rather low carbon isotopic ratios.
These stars have therefore already experienced the extra-mixing
episode, suggesting that the M4 bump could be slightly
fainter (by about 0.1~mag.).

On the other hand, our determinations for M4 stars
can be compared to the work of Shetrone (\cite{Shet}). 
First of all, it can be seen that 
the carbon isotopic ratios in both studies are very similar. 
In addition, we observed two of the M4 RGB stars studied by Shetrone:
the star M4\#304 (named M4\#4511 by Shetrone) and the star 
M4\#456 (identified as M4\#4507 by him). For the first one,
we measure \Crat=4$^{+1}_{-1}$, which is in very good agreement within the
error bars with Shetrone's \Crat=5$^{+2}_{-2}$ determination.
In contrary, we derived \Crat=7$^{+5}_{-2}$ for M4\#456, whereas Shetrone
found \Crat$>$15. We checked this discrepancy carefully
and are fairly confident of our  \Crat \ determination
since (i) the SNR of our spectrum appears to be better than
Shetrone's (see his Fig.~1) and (ii) M4\#456 is
slightly more luminous than the bump, so that 
it should have already experienced
the extra-mixing episode as confirmed by our rather low
estimate of its carbon isotopic ratio. Nevertheless, the same star is
slightly fainter than Shetrone's M4 bump estimate. This disagreement 
could be explained by the uncertainties in the M4 distance modulus, 
which is needed to
calculate the absolute magnitude values that he uses. 
Finally, 
Shetrone proposes that a small decline with
increasing luminosity (from just above the bump up to the tip)
was present in his data, suggesting a continued mixing.
The rather small statistics per individual GC, together with
the reported uncertainties on \Crat,  do not allow us to
confidently explore the existence of this trend.

Our conclusions can also be discussed with respect to the results of 
Gratton et al. (\cite{Grat}) who derived 
\Crat $\sim$ 6$^{+2}_{-1}$ up to 8$^{+2}_{-2}$
in nine field upper-RGB stars with metallicity
lower than -2~dex. Although some of their
estimates agree with ours, others
seem to be higher than the CNO cycle equilibrium value.
This discrepancy could result from the uncertainty
on the precise location of the onset of the extra-mixing episode
(i.e. the bump for GC stars) for field stars.
Indeed, the maximum inward penetration of the convective envelope
occurs at higher luminosities for higher stellar masses or lower metallicity.
In the Gratton et al. sample with stars with different
masses and metallicities, this penetration therefore occurs
at different luminosities. 
Thus, these RGB stars
with slightly higher carbon isotopic ratios could still be experiencing
their extra-mixing episode (in a GC, they would be found at or just
above the bump).

Finally, it has to be pointed out that Carretta et al. (\cite{Carr})
were the first to report determinations of the carbon isotopic ratio
in GC subgiant stars. They surprisingly found rather low \Crat \ ratios
(mean value around 8) for these not very evolved stars that have still
not experienced the 1DUP. They convincingly interpret their results
by invoking the pollution of the analysed stars by previously evolved
intermediate-mass AGB stars. If confirmed, this would reject
the commonly admitted assumption that, before the bump, the carbon
isotopic ratios in GC stars are still high. In consequence, the extra-mixing 
would not need to be as efficient as proposed by theoretical scenarios.

On the other hand, our derived carbon isotopic ratios for low-mass very metal
poor RGB stars are in very good agreement with extra-mixing  
estimates. For instance, Charbonnel (1995; see also
Denissenkov \& Vandenberg \cite{Den2}) proposes that rotation
induced extra-mixing leads to a decrease of \Crat \ down to 3.5
for stars with 0.8 M\sun \ and Z=$10^{-4}$ (i.e. close to
the properties of our targets).
In such low-metallicity stars, 
the mean molecular barrier, which inhibits  the extra-mixing, would 
be  much lower. That already leads to a very low \Crat \ ratio
just above the bump luminosity as observed in our work.
However, although the agreement appears very convincing,
these models are based on a parametric approach.
More physically realistic simulations of rotating low-mass stars have been
performed recently by Palacios et al. (\cite{Ana})
and show that self-consistent models, where
the rotational transport is treated within the
stellar evolution code, lead to chemical
variations that are too small at the stellar surface.
Open questions therefore still remain in the modelling
of rotating, evolved low-mass stars to reproduce
the evolution of the observed chemical abundances on their surface.

In summary, we therefore confirm that the extra-mixing scenario is 
efficient in any stellar system: field stars (Gratton et al., \cite{Grat}),
low (Shetrone, \cite{Shet}), and very-low (this work) metallicity GC.
Stars brighter than the RGB luminosity function bump
already show the signatures of that mixing.
Furthermore, within the uncertainties, the extra-mixing efficiency is found 
to be 
independent of the clusters metal content.

\begin{acknowledgements}
      We thank the anonymous referee for constructive remarks.
      We acknowledge the MARCS collaboration for its continuous efforts
      in computing stellar atmosphere models. B. Plez is also thanked 
      for providing molecular line lists and tools for computing stellar
      spectra. The VALD database was used when building the atomic lines.
      Y. Momany is acknowledged for his help on M30 astrometry.
      A. Recio-Blanco thanks the European Space Agency for  
      financial support through its post-doctoral fellowship programme.
      Many thanks to Magali and Benjamin for their careful reading of
      the manuscript!
\end{acknowledgements}


\begin{thebibliography}{}
  \bibitem[1999]{Alonso1} Alonso, A., Arribas, S., Martínez-Roger, C, 1999
                          \aaps, 140, 261
  \bibitem[2001]{Alonso2} Alonso, A., Arribas, S., Martínez-Roger, C, 1999
                          \aap, 376, 1039
  \bibitem[2005]{martin} Asplund, M., Grevesse, N., Sauval, A.J., et al.  
           2005, \aap, 431, 693
  \bibitem[1998]{Plez} Alvarez, R. \& Plez, B. 1998, \aap, 330, 1109
  \bibitem[1988]{Bess} Bessel, M.S., Brett, J.M., 1988, PASP 100, 1134
  \bibitem[1989]{Brown} Brown. J.A., Wallerstein, G. 1989, \aj, 98, 1643
  \bibitem[1989]{Card} Cardelli, J.A., Clayton, G.C., Mathis, J.S., 1989,
                       \apj, 345, 245
  \bibitem[2001]{Carp} Carpenter, J.M., 2001, AJ 121, 2851
  \bibitem[2005]{Carr} Carretta, E., Gratton, R.G., Lucatello, S., et al.,
                       2005, A\&A 433, 597
  \bibitem[1994]{Charb} Charbonnel, C. 1994, \aap,   282,  811
  \bibitem[1995]{Charb5} Charbonnel, C. 1995, \apj,   453,  L41
                         \aap,  336, 915
  \bibitem[1998]{Charb3}  Charbonnel, C., Brown, J.A., Wallerstein, G.
                1998, \aap, 332, 204
  \bibitem[2003]{Charb4}  Charbonnel, C., 2003, in {\it CNO in the Universe},
ASP Conf. Ser. 304, p.303, Charbonnel-Schaerer (Eds.)
  \bibitem[2002]{Cho}  Cho, D.-H., Lee, S.-G.,  2002, \aj, 124, 977
  \bibitem[1990]{Cud}  Cudworth, K.M., Rees, R., 1990, \aj, 99, 1491
  \bibitem[2004]{Den} Denissenkov, P.A., Herwig, F. 2004, \apj, 612, 1081
  \bibitem[2003]{Den2} Denissenkov, P.A., VandenBerg, D.A. 2003, \apj, 539, 509
  \bibitem[2000]{Ferr} Ferraro, F.R., Montegriffo, P., Origlia, L., 
                       Fusi Pecci, F. 2000, \aj, 119, 1282 
  \bibitem[1994]{CO} Goorvitch, D. \& Chackerian, C. 1994, \apjs, 91, 483.
  \bibitem[2000]{Grat} Gratton, R.G., Sneden, C., Carretta, E., et al. 2000, 
                       \aap, 354, 169
  \bibitem[1998]{GS} Grevesse, N. \& Sauval, A.J. 1998, Space Science 
           Reviews, 85, 161
  \bibitem[1974]{griffin} Griffin, R., 1974, \mnras, 167, 645
  \bibitem[2002]{Gus1} Gustafsson, B., Edvardsson, B., Eriksson, K., et al. 
           2002, ASP Conf. Ser. Vol. 288, (I.Hubeny, D.Mihalas, 
           K.Werner eds.), p. 331
  \bibitem[2006]{Gus2} Gustafsson, B., Edvardsson, B., Eriksson, K., 
           J\"orgensen, U.G., Nordlund, \AA, Plez B., 2006, in preparation
  \bibitem[1996]{Harris} Harris, W.E., 1996, \aj, 112, 1487
  \bibitem[1995]{hinkle} Hinkle, K., Wallace, L., Livingston, W., 1995,
           \pasp, 107, 1042.
  \bibitem[1964]{Iben} Iben, I.Jr.  1964, \apj, 140 1631
  \bibitem[1968]{Iben2} Iben, I.Jr.  1968, Nature 220, 143
  \bibitem[1985]{King} King, C.R., Da Costa, G.S., Demarque, P. 1985,
           \apj, 299, 674
  \bibitem[2003]{Kraft} Kraft, R.P., Ivans, I.I., 2003, \pasp, 115, 143
  \bibitem[1999]{Vald} Kupka, F., Piskunov, N.E., Ryabchikova, T.A.,  et al.
           1999, \aaps, 138, 119
  \bibitem[2004]{Yazan} Momany, Y., Bedin, L. R., Cassisi, S., et al. 2004, \aap, 420, 605
  \bibitem[2006]{Ana} Palacios, A., Charbonnel, C., Talon, S. et al., 2006,
                 \aap, 453, 261
  \bibitem[1993]{Peter} Peterson, R.C., Dalle Ore, C.M., Kurucz, R.L., 
           1993, \apj, 404, 333   
  \bibitem[2004]{Santi} Pietrinferni A., Cassisi S., Salaris M. and Castelli F., 2004, \apj, 612, 168
  \bibitem[1996]{Pila} Pilachowski, C.A., Sneden, C., Kraft, R.P., 1996,
                       \aj, 111, 1689
  \bibitem[2000a]{Ros1} Rosenberg, A., Piotto, G., Saviane, I., Aparicio, A., 2000a, A\&AS, 144, 5
  \bibitem[2000b]{Ros2} Rosenberg, A.; Aparicio, A.; Saviane, I.; Piotto, G., 2000b, A\&AS, 145, 451
  \bibitem[2003]{Shet} Shetrone, M.D., 2003, \apj, 585, L45
  \bibitem[2006]{2mass} Skrutskie, M.F., Cutri, R.M., Stiening, R., 2006, \aj, 131, 1163
  \bibitem[1989]{smith} Smith, V.V., Suntzeff, N.B., 1989, \aj, 97.1699  
  \bibitem[2000]{smith2} Smith, V.V., Hinkle, K.H.,  Cunha, K. et al. 2002, 
                   \aj, 124, 3241
  \bibitem[2004]{Val} Valenti, E., Ferraro, F.R., Origlia, L., 2004, 
                      \mnras, 354, 815 
  \bibitem[2006]{Weiss} Weiss, A., 2006, in {\it Chemical abundances and
      mixing in stars in the Milky Way and its satellites}, p.298, 
      ESO astrophysics symposia, Randich-Pasquini (Eds.), Springer
\end{thebibliography}
\end{document}